\documentclass[final]{svjour3}
\usepackage{graphicx}
\usepackage{rotating}
\usepackage{amssymb}
\usepackage{mathptmx}
\usepackage[numbers]{natbib}
\usepackage{lipsum}
\usepackage{footmisc}
\usepackage{color}
\makeatletter
\journalname{Journal of Low Temperature Physics}

\bibpunct{}{}{,}{s}{}{,}
\begin{document}
\newcommand{\hdblarrow}{H\makebox[0.9ex][l]{$\downdownarrows$}-}


\title{Development of Multi-Chroic MKIDs for Next-Generation CMB
  Polarization Studies}


\author{
B.~R.~Johnson$^\mathrm{a}$ \and
D.~Flanigan$^\mathrm{a}$ \and
M.~H.~Abitbol$^\mathrm{a}$ \and
P.~A.~R.~Ade$^\mathrm{b}$ \and
S.~Bryan$^\mathrm{c}$ \and
H.-M.~Cho$^\mathrm{g}$ \and
R.~Datta$^\mathrm{e,h}$ \and
P.~Day$^\mathrm{f}$ \and
S.~Doyle$^\mathrm{b}$ \and
K.~Irwin$^\mathrm{d,g}$ \and
G.~Jones$^\mathrm{a}$ \and
D.~Li$^\mathrm{g}$ \and
P.~Mauskopf$^\mathrm{c}$ \and
H.~McCarrick$^\mathrm{a}$ \and
J.~McMahon$^\mathrm{e}$ \and
A.~Miller$^\mathrm{a}$ \and
G.~Pisano$^\mathrm{b}$ \and
Y.~Song$^\mathrm{d}$ \and
H.~Surdi$^\mathrm{c}$ \and
C.~Tucker$^\mathrm{b}$
}

\institute{
a) Department of Physics, Columbia University, New York, NY, 10027, USA; \\
b) School of Physics \& Astronomy, Cardiff University, Cardiff, CF243AA, UK; \\
c) School of Earth and Space Exploration, Arizona State University, Tempe, AZ, 85287, USA; \\
d) Department of Physics, Stanford University, Stanford, CA, 94305-4085, USA; \\
e) Department of Physics, University of Michigan, Ann Arbor, MI, 48103, USA; \\
f) NASA, Jet Propulsion Lab, Pasadena, CA, 91109, USA; \\
g) SLAC National Accelerator Laboratory, Menlo Park, CA 94025, USA \\
h) NASA Goddard Space Flight Center, Mail Code 665, Greenbelt, MD 20771, USA \\
\email{bradley.johnson@columbia.edu}
}


\maketitle


\begin{abstract}

We report on the status of an ongoing effort to develop arrays of
horn-coupled, polarization-sensitive microwave kinetic inductance
detectors (MKIDs) that are each sensitive to two spectral bands
between 125 and 280~GHz.
These multi-chroic MKID arrays are tailored for next-generation,
large-detector-count experiments that are being designed to
simultaneously characterize the polarization properties of both the
cosmic microwave background (CMB) and Galactic dust emission.
We present our device design and describe laboratory-based measurement
results from two 23-element prototype arrays.
From dark measurements of our first engineering array we demonstrated
a multiplexing factor of 92, showed the resonators respond to bath
temperature changes as expected, and found that the fabrication yield
was 100\%.
From our first optically-loaded array we found the MKIDs respond to
millimeter-wave pulses; additional optical characterization
measurements are ongoing.
We end by discussing our plans for scaling up this technology to
kilo-pixel arrays over the next two years.


\end{abstract}




\begin{figure}[t]
\centering
\includegraphics[width=\textwidth]{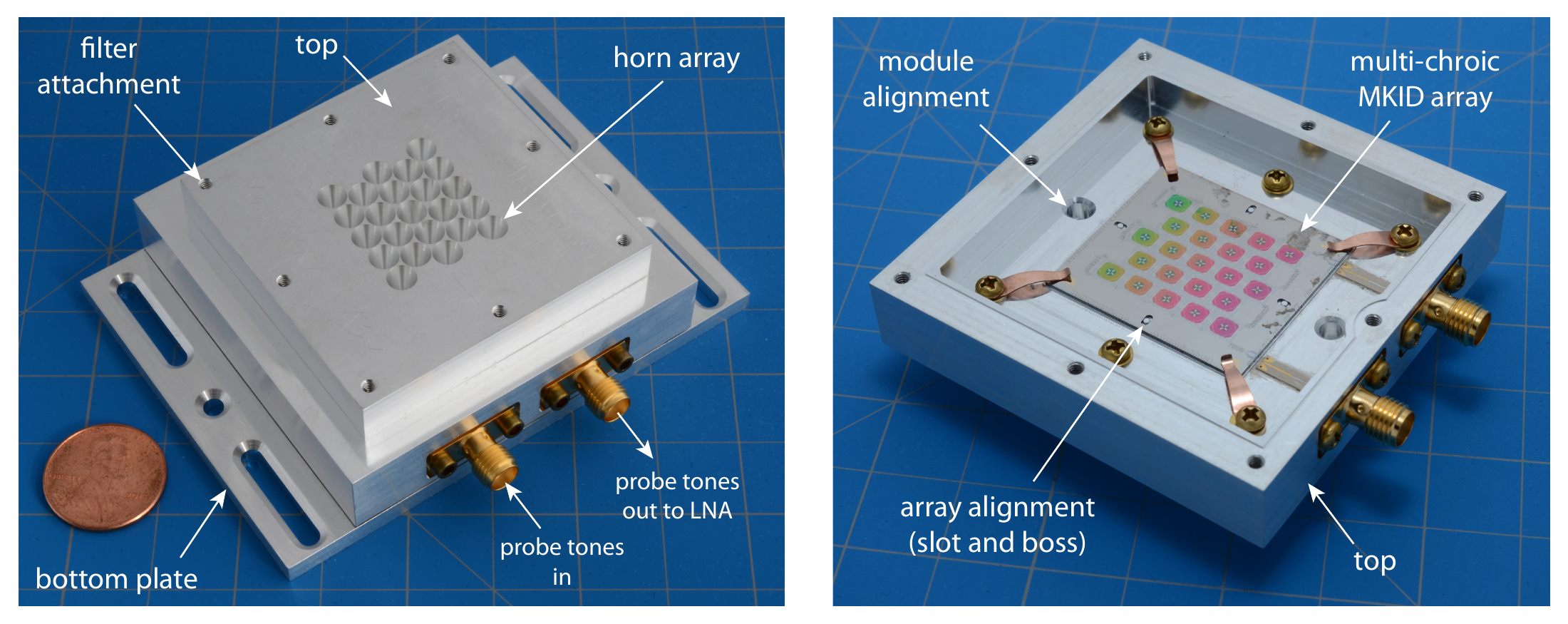}
\caption{
\textbf{Left:} Photograph of the 23-element multi-chroic MKID module
that is described in Section~\ref{sec:methods}.
The module consists of an aluminum enclosure and an MKID array, which
is mounted inside.
\textbf{Right:} The top of the enclosure contains 23 conical horns and
the detector array, which has a total of 92 MKIDs.
In this photo, the entrance apertures of the horns are facing down and
the bottom plate is removed.
The alignment of the MKID array is controlled using four bosses that
are machined in the aluminum enclosure and four slots that are etched
in the silicon with a deep-reactive ion etch (DRIE) process.
This configuration keeps the array aligned while allowing for
differential thermal contraction.
The slots are nominally 10~$\mu$m wider than the bosses, and this gap
sets the lateral alignment accuracy for the MKID array.
All of the MKIDs are read out with a single transmission line, so the
multiplexing factor for this detector system is 92.
}
\vspace{-0.5\baselineskip}
\label{fig:module}
\end{figure}


\vspace{-\baselineskip}
\section{Introduction}
\vspace{-0.4\baselineskip}
\label{sec:introduction}


In this paper we report on the status of an ongoing effort to develop
arrays of polarization-sensitive, microwave kinetic inductance
detectors (MKIDs) for future cosmic microwave background (CMB)
studies\cite{flanigan_2018,johnson_2016}.
The array elements we are developing are each sensitive to both
polarizations in two spectral bands, so there are four MKIDs per array
element.
These multi-chroic MKIDs are designed to help separate foreground
signals from CMB signals.
One of the spectral bands is for detecting the CMB, so it is centered
on 150~GHz, which is near the peak of the CMB blackbody spectrum.
The second spectral band is primarily for detecting Galactic dust
signals, so it is centered on 235~GHz, where Galactic dust emission is
brighter than the CMB.

MKIDs are planar superconducting GHz resonators.
The kinetic inductance and dissipation of the superconducting film
depend on the quasiparticle density.
When sufficiently energetic photons\footnote[2]{If $\nu > 2 \Delta/h
  \cong 74~\mathrm{GHz} \times (T_c/1~\mathrm{K})$, the photons will
  break Cooper pairs.  Here, $\Delta$ is the superconducting gap, $h$
  is Planck's constant, and $T_c$ is the superconducting transition
  temperature of the film.} are absorbed by the MKID, Cooper pairs
break, causing an increase in the quasiparticle density, a decrease in
the resonant frequency, and a decrease in the resonator quality
factor\cite{day_2003}.
These changes can be detected by monitoring the amplitude and phase of
a probe tone that drives the resonator at its nominal resonant
frequency.
Each MKID resonator is given a unique resonant frequency, so hundreds
to thousands of detectors in an array can be read out on a single
transmission line.
The detector architecture we describe in this paper is designed to be
well suited for next-generation CMB studies that require on the order
of five hundred thousand detectors\cite{cmb-s4_2017, cmb-s4_2016}.

Our development program is currently focused on building 23-element
prototype multi-chroic MKID modules and testing them in the
laboratory.
We present our module design in Section~\ref{sec:methods}. 
To date, we have fabricated and tested one module and one engineering
array.
Our measurement results are presented in Section~\ref{sec:results}.
Finally, over the next two years we will scale up our module design
and build a kilo-pixel array.
These future plans are discussed in Section~\ref{sec:discussion}.


\begin{figure}[t]
\centering
\includegraphics[width=\textwidth]{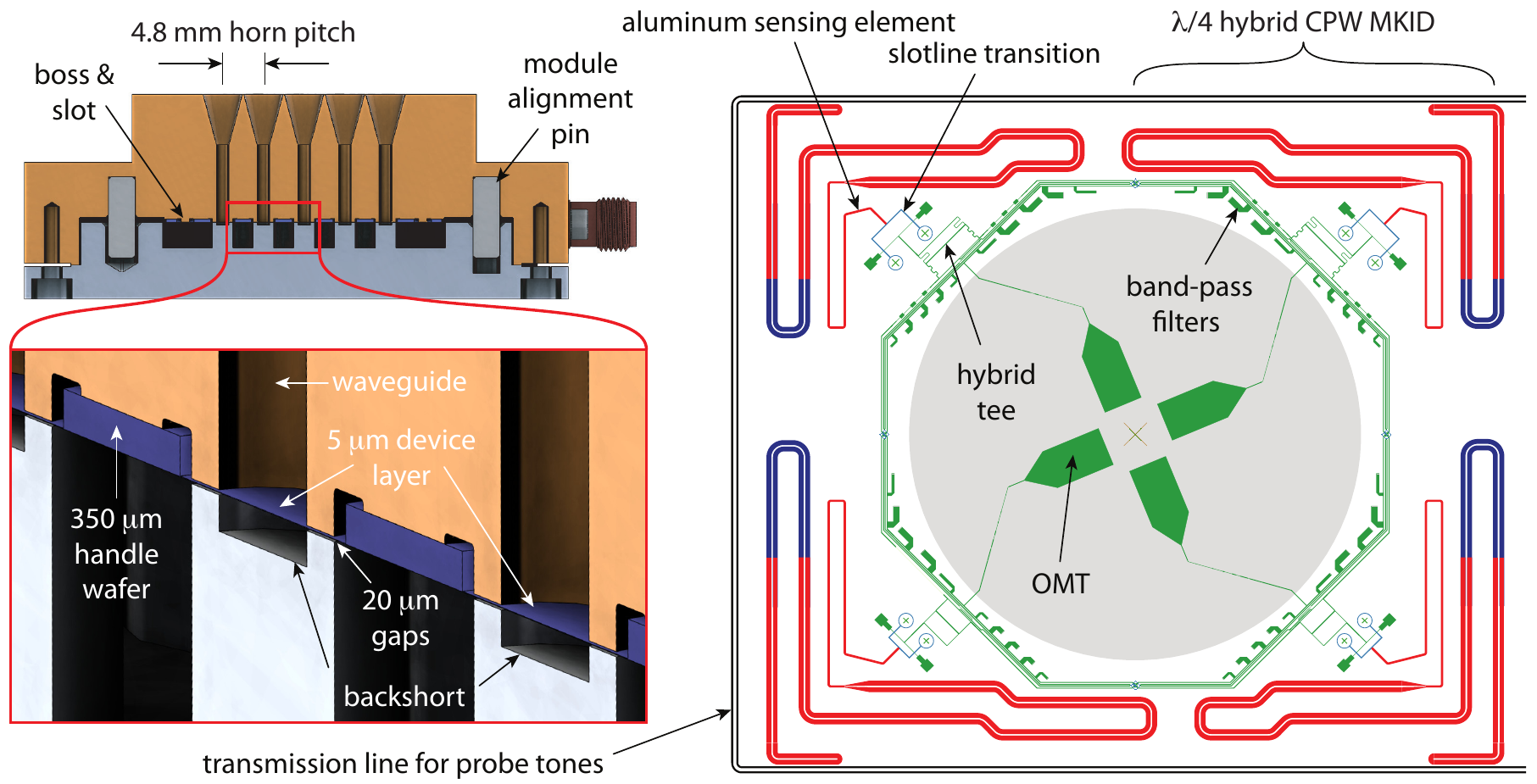}
\caption{
\textbf{Left:} A cross-sectional view of the multi-chroic MKID module.
Each conical horn feeds a cylindrical waveguide, which in turn, feeds
an array element.
The array is fabricated on a silicon-on-insulator (SOI) wafer.
Backshorts are used to optimize the millimeter-wave coupling.
\textbf{Right:} A scale drawing of one array element.
The cylindrical waveguide feeds the probe pairs in the orthomode
transducer (OMT).
Sky signals are filtered with a microstrip band-pass and then coupled
to the coplanar waveguide (CPW) MKID with a novel slotline
transition\cite{surdi_2016}.
Each array element has four MKIDs -- two polarizations in each of the
two spectral bands.
}
\vspace{-0.5\baselineskip}
\label{fig:pixel_design}
\end{figure}


\vspace{-\baselineskip}
\section{Methods}
\vspace{-0.4\baselineskip}
\label{sec:methods}



The 23-element prototype modules we are currently building are
composed of an MKID array fabricated on a silicon-on-insulator (SOI)
wafer that is mounted inside a two-piece aluminum enclosure.
The top of the aluminum enclosure is a monolithic horn plate that also
serves as the mounting surface for the MKID array.
The bottom plate, which closes the module, contains backshorts, which
are used to optimize the millimeter-wave coupling.
The aluminum enclosure is machined with a high-precision Kern 44
computer numeric control (CNC) mill, so the top and the bottom of the
enclosure are aligned to within a few microns.
A photograph of the module is shown in Figure~\ref{fig:module}, and
the array-element design is shown in Figure~\ref{fig:pixel_design}.

A conical horn with a 4.66~mm diameter aperture and a 15~deg flare
angle is used to feed each array element in the module.
We chose to use conical horns in the prototype module for convenience,
and we will switch to profiled horns in the future\cite{kittara_2008}.
Light emerging from the cylindrical waveguide at the back of the
conical horn is coupled to a broadband planar orthomode transducer
(OMT).
The waveguide diameter is 1.49~mm and we made it approximately 9~mm
long to ensure evanescent low-frequency modes do not reach the
detectors.
The OMT separates the two polarizations of the incoming
millimeter-wave signal using two waveguide probe
pairs\cite{datta_2016a}.
A choke around the exit aperture of the cylindrical waveguide improves
the waveguide/OMT coupling efficiency by minimizing the fields that
leak laterally into the module along the array.

\begin{figure}[t]
\centering
\includegraphics[width=0.7\textwidth]{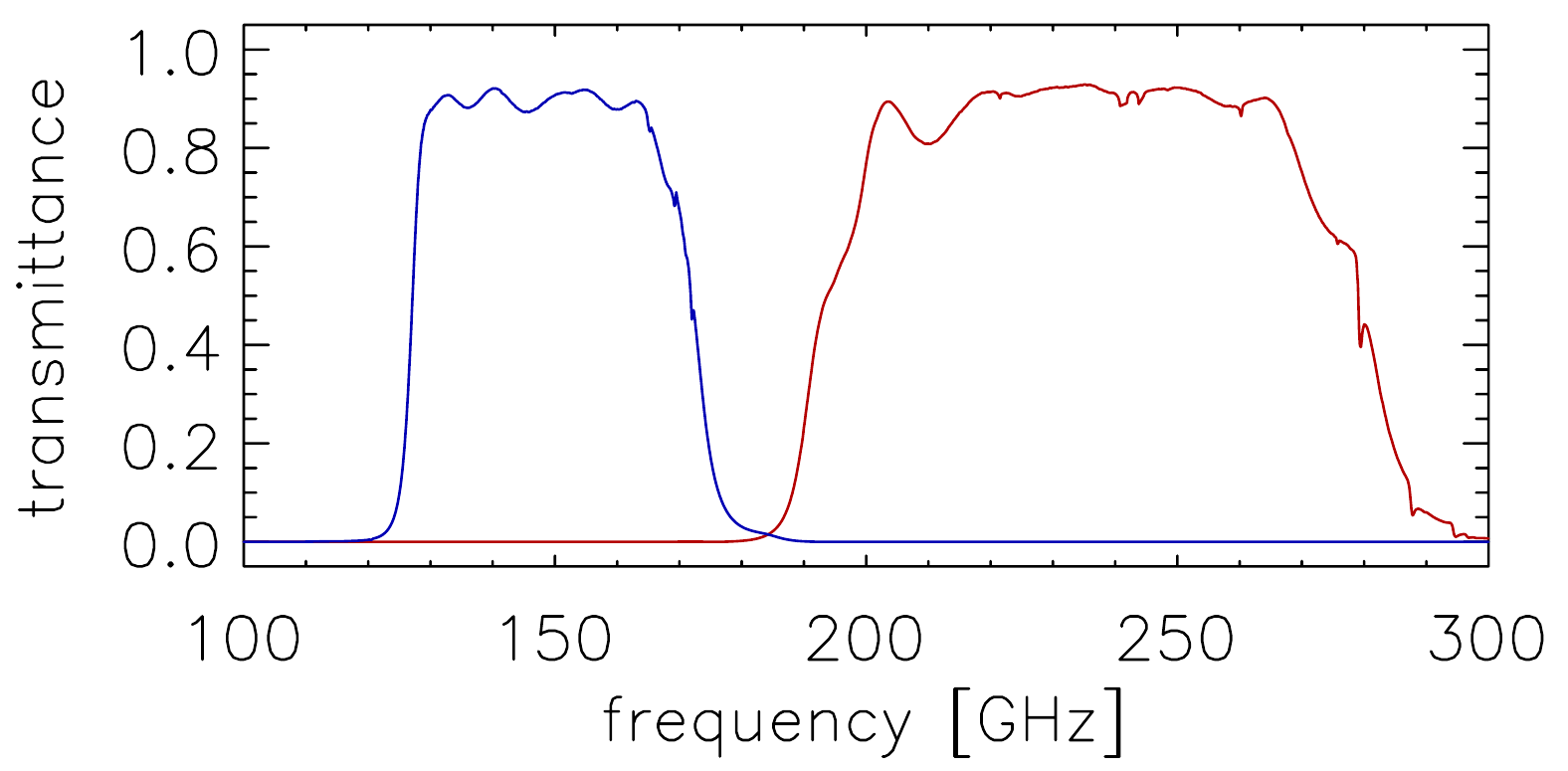}
\caption{
Simulated spectral bands in each of the multi-chroic MKID array
elements\cite{johnson_2016,datta_2016b,surdi_2016}.
The electromagnetic simulations that were used to compute these curves
start at the CPW output of the OMT, include all of the millimeter-wave
circuit components, and end in the absorbing element of the MKID (see
Section~\ref{sec:methods}).
The expected absorption efficiency is approximately 90\% across both
the 150~GHz and the 235~GHz spectral bands.
A metal-mesh low-pass filter with an edge near 360~GHz will be used to
reject high-frequency, out-of-band radiation.
The metal-mesh filter will be mounted outside the module in front of
the horn apertures (see Figure~\ref{fig:module}).
The transmittance of this filter and any frequency-dependent effects
from the waveguide-to-OMT coupling are not included in these
simulations.
}
\vspace{-0.5\baselineskip}
\label{fig:spectral_bands}
\end{figure}

The two spectral bands (125~to~170~GHz and 190~to~280~GHz) are defined
by five-pole resonant-stub microstrip band-pass filters (see
Figure~\ref{fig:spectral_bands}).
The output of each waveguide probe is coplanar waveguide (CPW).
Therefore, a broadband CPW-to-microstrip transition is first used to
connect the waveguide probes to the filter circuit.
This transition is composed of seven alternating sections of CPW and
microstrip.
The signals from each probe pair within a single spectral band are
then combined using a 180$^{\circ}$ hybrid.
The dimensions of all of these millimeter-wave circuit components are
given in the literature\cite{datta_2016b}.
The TE$_{11}$ waveguide mode ultimately couples to the difference port
on the 180$^{\circ}$ hybrid, while higher-order modes couple to the
sum port\cite{johnson_2016,datta_2016a}.
Therefore, to ensure single-moded performance over the required
bandwidth ratio of 2.25:1, signals at the sum port of the
180$^{\circ}$ hybrid are routed to a termination resistor and
discarded, while signals at the difference port are sent to the MKID
using a broadband coupling circuit\cite{surdi_2016} (described below).


The MKIDs are $\lambda/4$ resonators composed of two sections of CPW
in series.
The center line of the resonator is electrically grounded at one end
and capacitively coupled at the other end to the transmission line
used to read out the array.
The section of CPW at the grounded end of the resonator is the
absorptive section, and the center line here is made from aluminum.
Our thin-film aluminum has $T_c$ = 1.4~K, so photons with frequencies
greater than approximately 100~GHz will be energetic enough to break
Cooper pairs in the aluminum.
It is desirable to make the volume of this absorber as small as
possible to increase the responsivity.
As a starting point, we chose the dimensions of our absorbing element
to be 4~$\mu$m wide, 40~nm thick, 2.1~mm long for the 150~GHz spectral
band and 2.7~mm long for the 235~GHz spectral band.
The absorbing element is longer for the 235~GHz spectral band because
we anticipate there will be more loading from the sky, so the volume
needs to be larger.
Simulations show that with these lengths, approximately 90\% of the
millimeter-wave power from the sky will be absorbed in the aluminum
and break Cooper pairs\cite{johnson_2016}.
The gaps between the aluminum center line and the ground plane are
5~$\mu$m wide.
The second section of center line is made from a niobium-over-aluminum
bilayer, which has $T_c$ = 8.3~K.
The photons in our two spectral bands will not be energetic enough to
break Cooper pairs in this section of the resonator.
This second section of CPW is used to tune the resonant frequency of
the MKID, so the length varies from detector to detector; the length
range is 8.8 to 10.4~mm.
The end of the resonator near the readout transmission line has the
largest electric fields and is therefore most susceptible to two-level
system (TLS) effects\cite{gao_2008a,gao_2008b}.
To minimize any TLS effects, the gap to the ground plane in the CPW
should be wide.
As a starting point, we made the niobium center line 10~$\mu$m wide
and the gap to the ground plane 30~$\mu$m wide.
This $\sim$10~mm total length of CPW produces resonant frequencies
near 3~GHz.
Hybrid CPW MKIDs like these have been shown to be photon-noise limited
over a wide range of millimeter and sub-millimeter wavelengths with
optical loading levels well under 1~pW\cite{janssen_2014,
  yates_2011}.

To couple the microstrip output of the 180$^{\circ}$ hybrid to the
MKID, the power is first evenly divided in-phase onto two microstrips,
each with twice the impedance of the microstrip from the 180$^{\circ}$
hybrid.
Each branch feeds a standard broadband microstrip-to-slotline
transition, where the slotline is formed in the niobium ground plane
that is common to the microstrip circuit and the MKID.
The two slotlines are then brought together and become the CPW gaps in
the aluminum section of the MKID.


\begin{figure}[t]
\centering
\includegraphics[width=\textwidth]{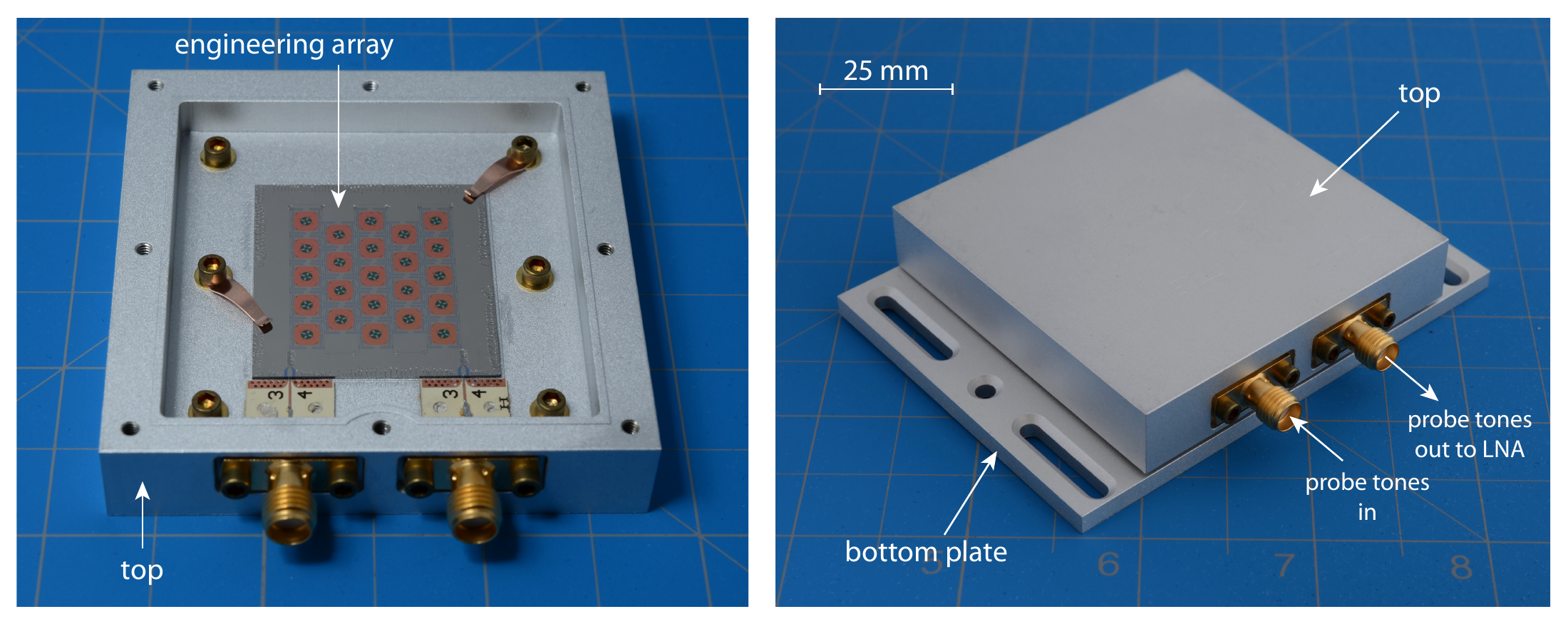}
\caption{
\textbf{Left:} Photograph of the engineering array mounted inside a
separate aluminum enclosure.
\textbf{Right:} This second enclosure does not have any conical horns,
so after the bottom plate is installed and the module is sealed, the
MKIDs are not externally illuminated.
Characterization measurements of the resonators in this engineering
array are presented in Section~\ref{sec:results}.
}
\vspace{-0.5\baselineskip}
\label{fig:dark_module}
\end{figure}



The MKID arrays are fabricated on SOI wafers 100~mm in diameter.
Each SOI wafer consists of a 5~$\mu$m thick float-zone silicon
($>$10~k$\Omega$\,cm resistivity) device layer and a 350~$\mu$m thick
silicon handle wafer held together by a 0.5~$\mu$m thick buried oxide
layer.
An aluminum/niobium bilayer is first deposited on the device layer.
The aluminum is 40~nm thick and the niobium is 200~nm thick.
This bilayer is patterned to produce the OMTs, the MKIDs, and the
ground plane.
A 350~nm thick film of silicon nitride (SiN$_x$) is deposited on top
of the bilayer, followed by a 400~nm thick niobium film.
The SiN$_x$ serves as the electrically insulating dielectric material
in the microstrip, and the niobium film is patterned to form the
microstrip circuit that includes the band-pass filters and the
180$^{\circ}$ hybrids.
Our design uses cross-unders\cite{duff_2016} in the microstrip circuit
rather than cross-overs, which decreases the number of required
fabrication steps.
A gold film is deposited and patterned on top of the SiN$_x$ to
construct the aforementioned termination resistor at the sum port of
the 180$^{\circ}$ hybrid.
The SiN$_x$ is removed near the MKIDs to reduce loss and two-level
system (TLS) noise\cite{zmuidzinas_2012}.
The bilayer niobium is removed from the $\sim$2~mm long sensing
section of the center line of the MKID leaving only the aluminum.
To improve the optical coupling and to minimize TLS noise, the thick
silicon handle wafer and the buried oxide layer underneath the OMT and
the high-field section of the MKID are removed using deep reactive ion
etching (DRIE).


\vspace{-\baselineskip}
\section{Results}
\vspace{-0.4\baselineskip}
\label{sec:results}


We first produced an engineering array on a monolithic 500~$\mu$m
thick float-zone silicon wafer to test a majority of the processing
steps (see Figure~\ref{fig:dark_module}).
This engineering array is not optimized for millimeter-wave coupling
because the substrate is too thick.
Therefore, we mounted it in a second aluminum enclosure with no horns,
chokes, or backshorts and ran some resonator tests where the MKIDs
were not illuminated.

\begin{figure}[t]
\centering
\includegraphics[height=0.145\textheight]{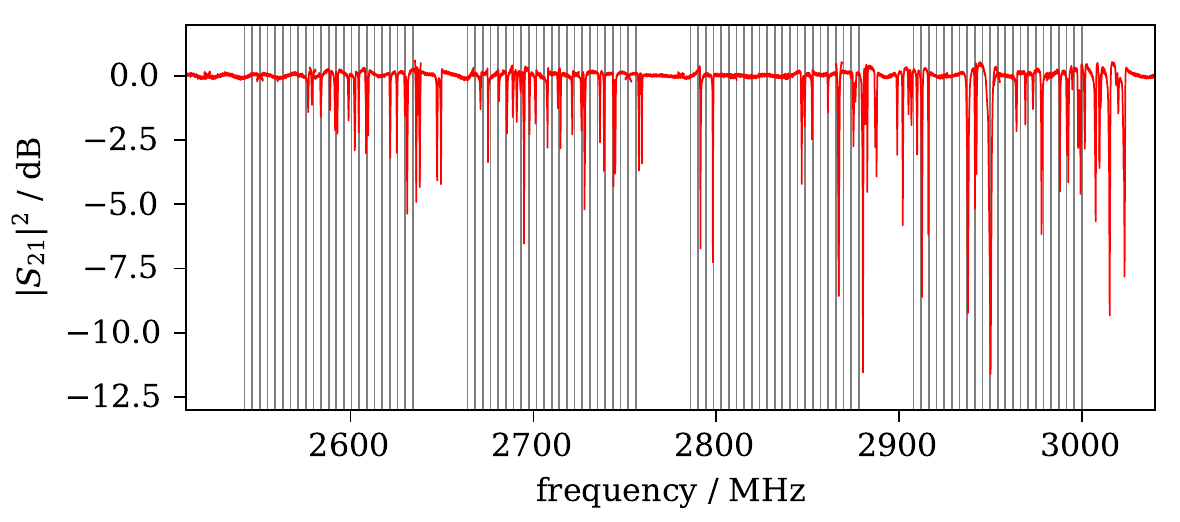}
\includegraphics[height=0.145\textheight]{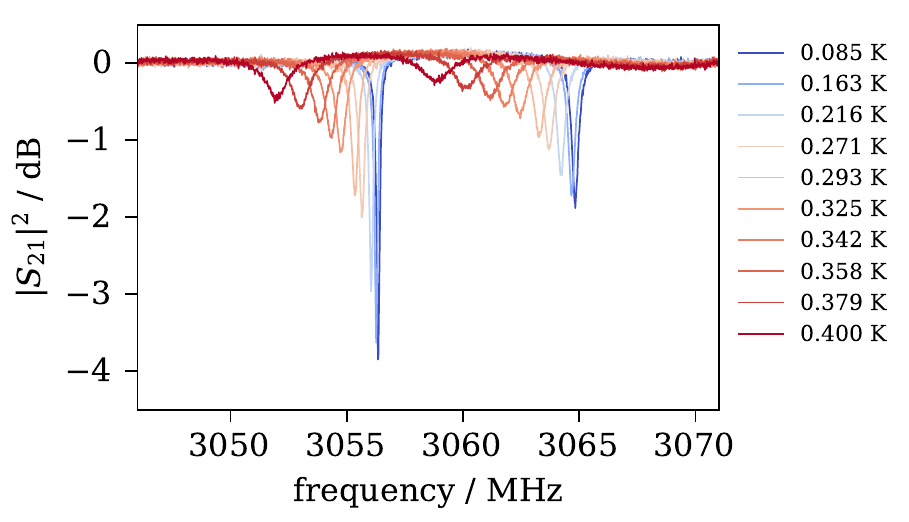}
\includegraphics[height=0.17\textheight]{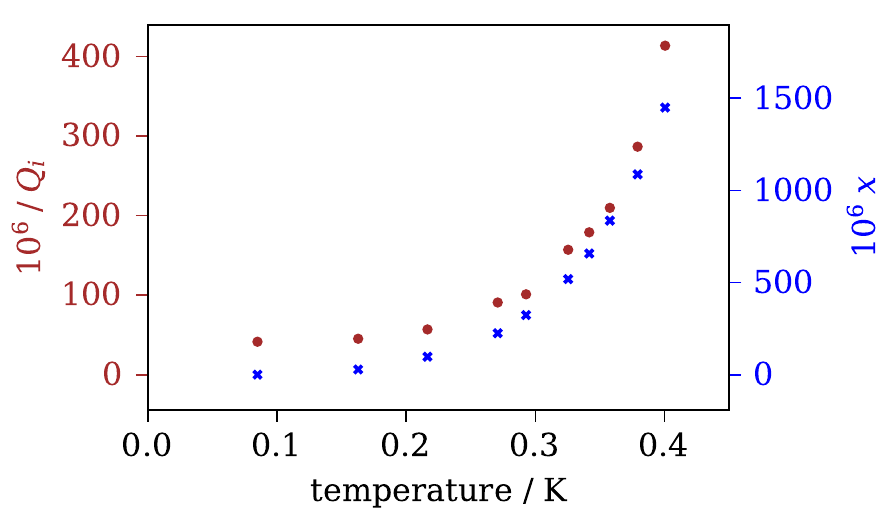}
\includegraphics[height=0.17\textheight]{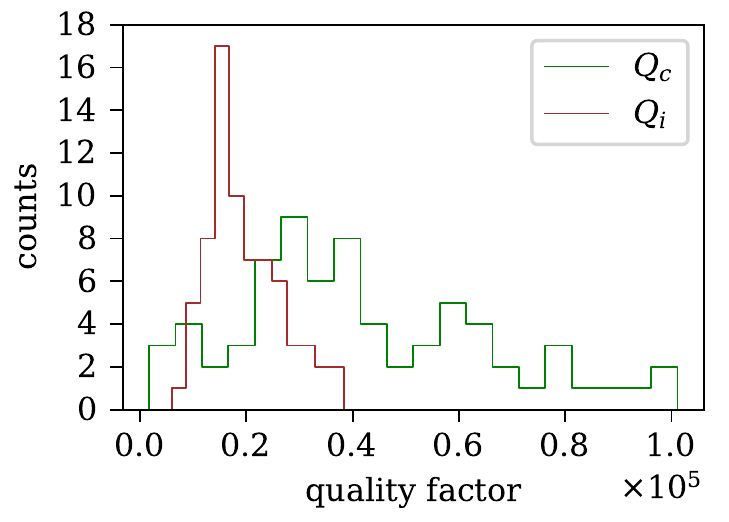}
\caption{
Measurements of the resonators in the engineering array (see
Figure~\ref{fig:dark_module}).
\textbf{Top Left:} Measured complex forward transmission ($S_{21}$) as
a function of frequency.
The vertical gray lines indicate the designed resonant frequencies.
For this measurement, the module temperature was 200~mK.
\textbf{Top Right:} Measured $S_{21}$ for two resonators as a function of
temperature.
The resonant frequency and quality factor decrease when the
temperature increases, as expected.
\textbf{Bottom Left:} Measured fractional frequency shift $x$ and
inverse internal quality factor $Q_i^{-1}$ as a function of module
temperature for a 2695~MHz resonator.
The internal quality factor of the resonator is related to $Q$ and
$Q_c$ as $Q^{-1} = Q_i^{-1} + Q_c^{-1}$.
\textbf{Bottom Right:} Histogram of $Q_i$ and $Q_c$ measured at 85~mK.
Data from 71 of the 92 resonators are included in the histogram.
Data from the additional 21 resonators require more careful fitting,
in part because many of them partially overlap in frequency, so we did
not include them in this initial study.
Future arrays will include lithographed connections between the ground
planes on opposite sides of the transmission line CPW, which should
reduce the scatter in resonance frequencies.
}
\vspace{-0.5\baselineskip}
\label{fig:data}
\end{figure}

In these tests, we drove the resonators with probe tones via the
readout transmission line, and we measured the amplitude and phase of
the emerging waveforms.
These tests were done over a range of module temperatures between
85~mK and 400~mK.
The details of the readout system we used are given in the
literature\cite{johnson_2016}.
By sweeping the frequency of the probe tone from 1.8 to 4.0~GHz, we
measured the complex forward transmission ($S_{21}$).
The $S_{21}$ data is shown in the top panels of Figure~\ref{fig:data}.
To characterize the resonances, data from each resonator was fit to
the equation\cite{khalil_2012}
\begin{equation}
\label{eq:s21}
S_{\mathrm{21}} = 1 - \frac{Q}{Q_c} \left(\frac{1}{1 + 2 j x Q}\right),
\end{equation}
where $Q_c$ is the complex coupling quality factor, $Q$ is the
resonator quality factor, $x = 1 - f/f_0$ is the fractional frequency
shift, $f$ is the variable resonant frequency, $f_0$ is the reference
resonant frequency, and we have omitted an overall phase term for
clarity.
This measurement revealed (i) the resonant frequencies of all 92
resonators in the array, so the MKID fabrication yield was 100\%, and
(ii) the resonant frequency and quality factor of the resonator
changed with bath temperature as expected.

\begin{figure}[t]
\centering
\includegraphics[width=0.7\textwidth]{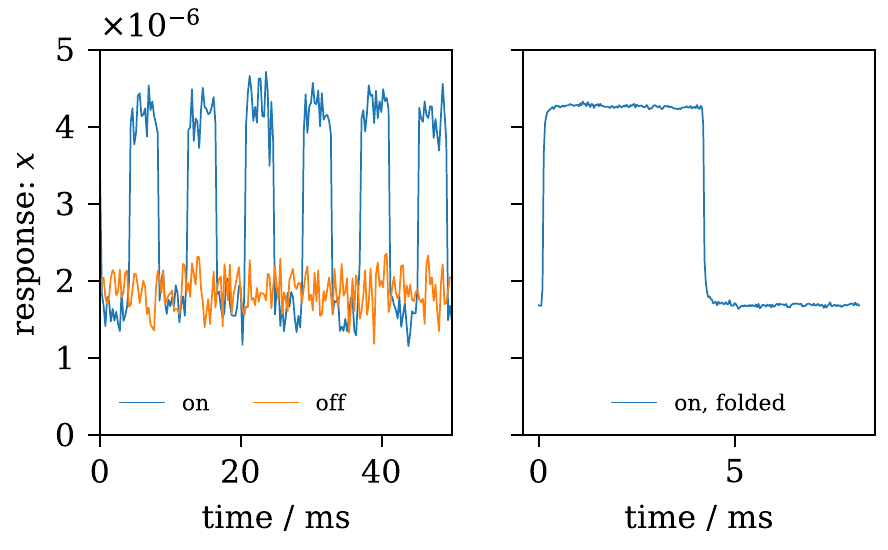}
\caption{
Millimeter-waves detected with our first multi-chroic MKID module (see
Figure~\ref{fig:module}).
The incoherent broadband (140 to 160~GHz) millimeter-wave pulses were
generated using an electronic
source\cite{mccarrick_2018,flanigan_2016}.
\textbf{Left:} Time-ordered data with the millimeter-wave pulses both
on and off.
\textbf{Right:} To clearly show the shape of the detected pulse, many
pulses have been folded down to one period by averaging samples
separated in time by an integer number of pulse periods.
}
\vspace{0\baselineskip}
\label{fig:photon_detection}
\end{figure}

Characterization measurements of our first optically-alive module (see
Figure~\ref{fig:module}) are underway\cite{flanigan_2018}.
These measurement results will be published in the future.
However, in Figure~\ref{fig:photon_detection} we show some preliminary
data: the MKIDs are responding to millimeter-waves generated with an
electronic source that is rapidly switched on and off with a PIN
diode.
For this measurement, we used a cryogenic test system that was
previously used for lumped-element kinetic inductance detector
studies, and the details of this test system are described in the
literature\cite{mccarrick_2018,flanigan_2016}.


\vspace{-\baselineskip}
\section{Discussion}
\vspace{-0.4\baselineskip}
\label{sec:discussion}

\begin{figure}[t]
\centering
\includegraphics[width=\textwidth]{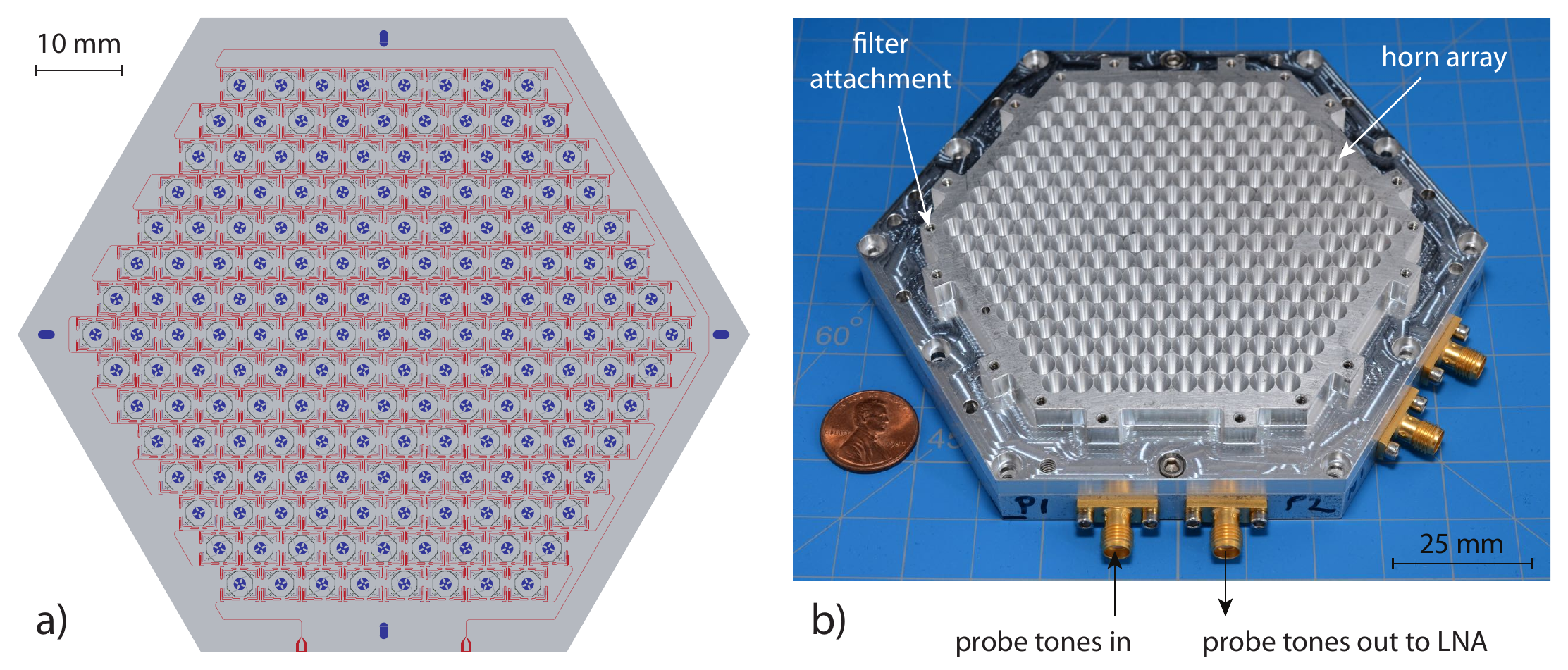}
\caption{
\textbf{Left:} Scale drawing of the 169-element multi-chroic MKID
array we are developing.
\textbf{Right:} Photograph of a hexagonal aluminum module.
This particular module has 271 horns and it was made for
lumped-element kinetic inductance
detectors\cite{mccarrick_2018}, but the multi-chroic
MKID module we will build will look similar.
}
\vspace{-0.5\baselineskip}
\label{fig:future}
\end{figure}

Over the next two years, we will develop a hexagonal multi-chroic MKID
module that contains 169 array elements (676 MKIDs).
Multiple modules can therefore be arranged to form a multi-kilo-pixel
array.
The array design and a concept enclosure are shown in
Figure~\ref{fig:future}.
For this new module we will switch to profiled horns and replace the
aluminum films in the array with aluminum-manganese (AlMn) films.
AlMn has a tunable $T_c$ that can be decreased below 1.4~K.
Reducing the sensor $T_c$ increases the average number of Cooper pairs
broken by each photon in our two spectral bands, and thus decreases
the relative contribution of recombination noise\cite{flanigan_2016}.
Members of our collaboration pioneered the use of AlMn as a material
for transition edge sensor (TES) bolometers\cite{li_2016}, and we
have already shown that AlMn works as a high-Q resonator
material\cite{jones_2016}.


\begin{acknowledgements}

This work is supported by NSF grants AST-1509211 and AST-1711160 for
Johnson; AST-1509078 and AST-1711242 for Mauskopf; AST-1506074 and
AST-1710624 for Irwin.
McCarrick is supported by a NASA Earth and Space Sciences Fellowship.
We thank the Xilinx University Program for donating the FPGA hardware
and software tools that were used in the readout system.

\end{acknowledgements}


\vfill


\pagebreak


\bibliographystyle{unsrt}
\bibliography{references}


\end{document}